%% file: OrderParameters.tex
\documentclass[showpacs,preprintnumbers,showkeys,twocolumn,prl,floatfix]{revtex4}
\usepackage{epsfig}

\begin{document}

\input{timeToday}

\title{Order Parameters, Broken Symmetry, and Topology}
\date{\timetoday}

\author{James P. Sethna}
\affiliation{
Laboratory of Applied Physics, Technical University of Denmark, %
DK-2800 Lyngby, DENMARK, and NORDITA, DK-2100 Copenhagen \O, DENMARK
and
Laboratory of Atomic and Solid State Physics (LASSP), Clark Hall,
Cornell University, Ithaca, NY 14853-2501, USA}

\begin{abstract}
We introduce the theoretical framework we use to study the bewildering
variety of phases in condensed--matter physics.  We emphasize the
importance of the breaking of symmetries, and develop the idea of an
order parameter through several examples.  We discuss elementary
excitations and the topological theory of defects.

\noindent
{\bf 1991 Lectures in Complex Systems,} Eds.\ L.~Nagel and D.~Stein,
Santa Fe Institute Studies in the Sciences of Complexity, Proc.\ Vol.\ XV,
Addison-Wesley, 1992.
\end{abstract}

\maketitle

As a kid in elementary school, I was taught that there were three states
of matter: solid, liquid, and gas.  The ancients thought that there
were four: earth, water, air, and fire, which was considered sheer
superstition.  In junior high, I remember reading a book called {\sl The Seven
States of Matter}.  At least one was ``plasma'', which made up stars and thus
most of the universe,\footnote{They hadn't heard of dark matter
back then.} and which sounded rather like fire to me.

\begin{figure}[thb]
\epsfxsize=2.5truein
\epsffile{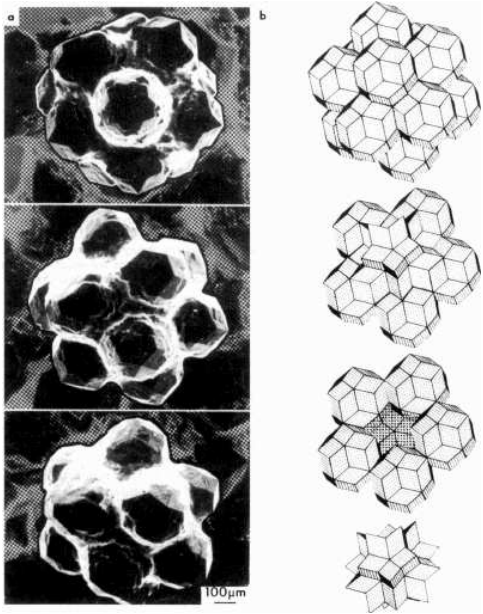}
\caption{{\bf Quasicrystals.}
Much of these two lectures will discuss the properties of crystals.
Crystals are surely the oldest known of the broken--symmetry phases of
matter, and remain the most beautiful illustrations.  It's amazing
that in the past few years, we've uncovered an entirely new class of
crystals.  Shown here is a photograph of a quasicrystalline metallic alloy,
with icosahedral symmetry.  Notice that the facets are pentagonal:
our old notions of crystals had to be completely revised to include
this type of symmetry.
}
\label{fig:Quasicrystal}
\end{figure}

The original three, by now, have become multitudes.  In important and precise
ways, magnets are a distinct form of matter.  Metals are different from
insulators.  Superconductors and superfluids are striking new states
of matter.  The liquid crystal in your wristwatch is one of a huge
family of different liquid crystalline states of matter\cite{deGennes}
(nematic, cholesteric, blue phase I, II, and blue fog, smectic A, B, C,
C$^*$, D, I, ...).  There are over 200 qualitatively different types of
crystals, not to mention the quasicrystals (figure~1).  There are
disordered states of matter like spin glasses, and states like the
fractional quantum hall effect with excitations of charge $e/3$ like quarks.
Particle physicists tell us that the vacuum we live
within has in the past been in quite different states: in the last vacuum
but one, there were four different kinds of light\cite{Coleman}
(mediated by what is now the photon, the W$^+$, the W$^-$, and the Z particle).
We'll discuss this more in lecture two.

When there were only three states of matter, we could learn about each one
and then turn back to learning long division.  Now that there are
multitudes, though, we've had to develop a system.  Our system is
constantly being extended and modified, because we keep finding new
phases which don't fit into the old frameworks.  It's amazing how the
500th new state of matter somehow screws up a system which worked
fine for the first 499.  Quasicrystals, the fractional quantum hall effect,
and spin glasses all really stretched our minds until (1)~we understood
why they behaved the way they did, and (2)~we understood how they fit
into the general framework.

In this lecture, I'm going to tell you the system.  In the
next three lectures, I'll discuss some gaps in the system: materials and
types of behavior which don't fit into the neat framework presented here.
I'll try to maximize the number of pictures and minimize the number of
formulas, but (particularly in lecture III) there are problems and ideas
that I don't understand well enough to explain simply.  Most of what
I tell you in this lecture is both true and important.  Much of what
is contained in the next three lectures represents my own pet ideas and
theories, and you should be warned not to take my messages there as
gospel.

The system consists of four basic steps.\cite{Mermin}  First, you must
identify
the broken symmetry.  Second, you must define an order parameter.  Third,
you are told to examine the elementary excitations.  Fourth, you
classify the topological defects.  Most of what I say I take from
Mermin\cite{Mermin}, Coleman\cite{Coleman}, and deGennes\cite{deGennes},
and I heartily recommend these excellent articles to my audience.
We take each step in turn.

\section{Identify the Broken Symmetry}

What is it which distinguishes the hundreds of different states of matter?
Why do we say that water and olive oil are in the same state (the liquid
phase), while we say aluminum and (magnetized) iron are in different states?
Through long experience, we've discovered that most phases differ in their
symmetry.\footnote{This is not to say that different phases always
differ by symmetries!  Liquids and gases have the same symmetry.  In fact,
one can go continuously from a liquid to a gas, by going first to high
pressures and then heating.  It is safe to say, though, that if
the two materials have different symmetries, they are different phases.}

\begin{figure}[thb]
\center{
\epsfxsize=1.25truein
\epsffile{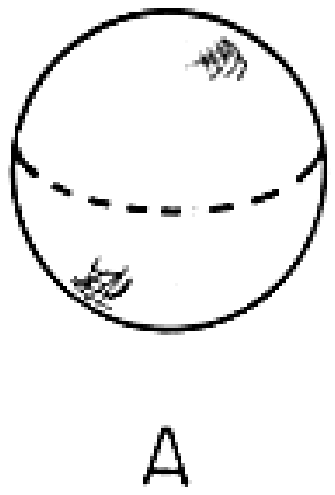}
\hskip 0.2truein
\epsfxsize=1.25truein
\epsffile{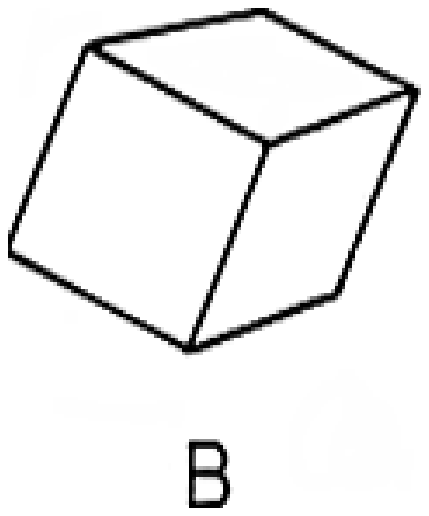}}
\caption{{\bf Which is more symmetric?}
The cube has many symmetries.  It can be rotated by $90^\circ$, $180^\circ$,
or $270^\circ$ about any of the three axes passing through the faces.
It can be rotated by $120^\circ$ or $240^\circ$ about the corners,
and by $180^\circ$ about an axis passing from the center through
any of the 12 edges. The sphere, though, can be rotated by {\it any} angle.
The sphere respects rotational invariance: all directions are equal.
The cube is an object which breaks rotational symmetry: once the cube
is there, some directions are more equal than others.
}
\label{fig:2}
\end{figure}

Consider figure~2, showing a cube and a sphere.  Which is more symmetric?
Clearly, the sphere has many more symmetries than the cube.  One can
rotate the cube by 90$^\circ$ in various directions and not change its
appearance, but one can rotate the sphere by any angle and keep it
unchanged.

\begin{figure}[thb]
\center{
\epsfxsize=1.25truein
\epsffile{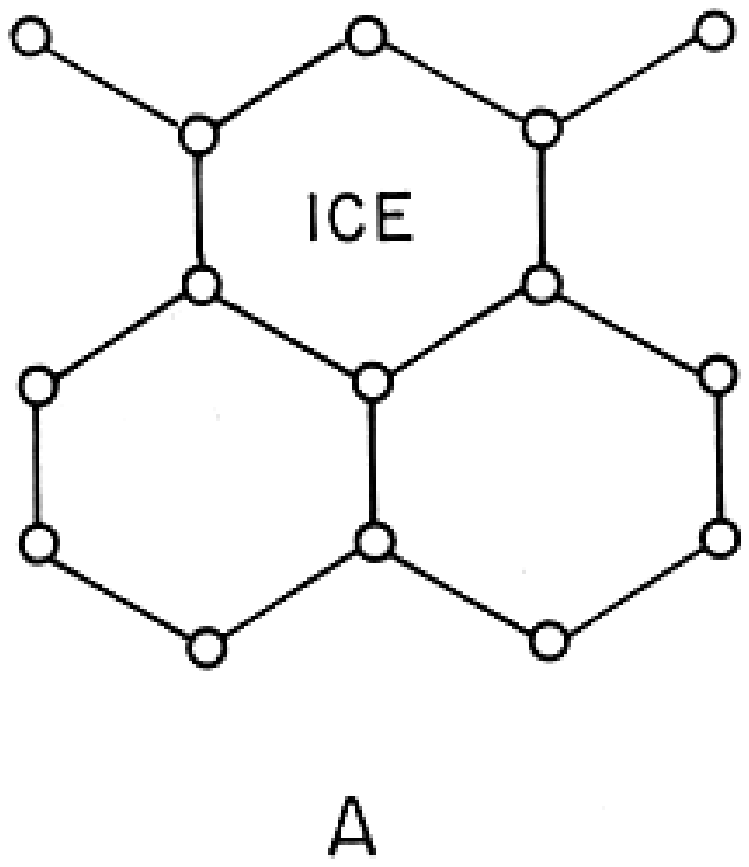}
\hskip 0.2truein
\epsfxsize=1.25truein
\epsffile{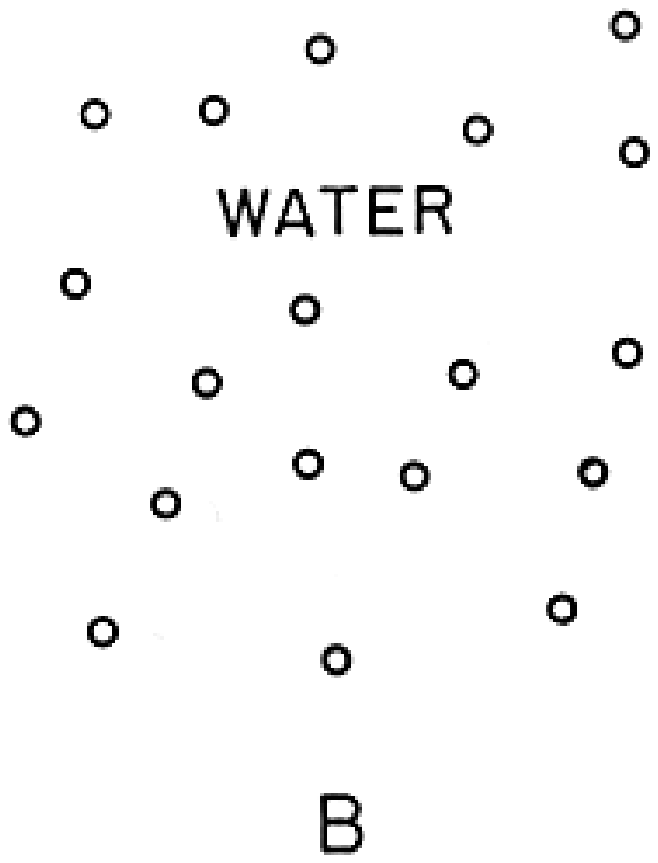}}
\caption{{\bf Which is more symmetric?}
At first glance, water seems to have much less symmetry than ice.
The picture of ``two--dimensional'' ice clearly breaks the rotational
invariance: it can be rotated only by $120^\circ$ or $240^\circ$.
It also breaks the translational invariance: the crystal can only
be shifted by certain special distances (whole number of lattice units).
The picture of water has no symmetry at all: the atoms are
jumbled together with no long--range pattern at all.  Water, though,
isn't a snapshot: it would be better to think of it as a combination
of all possible snapshots!  Water has a complete rotational and translational
symmetry: the pictures will look the same if the container is tipped or
shoved.
}
\label{fig:3}
\end{figure}

In figure~3, we see a 2-D schematic representation of ice and water.
Which state is more symmetric here?  Naively, the ice looks much more
symmetric: regular arrangements of atoms forming a lattice structure.
The water looks irregular and disorganized.  On the other hand, if
one rotated figure~3B by an arbitrary angle, it would still look like
water!  Ice has broken rotational symmetry: one can rotate figure~3A
only by multiples of 60$^\circ$.  It also has a broken translational
symmetry: it's easy to tell if the picture is shifted sideways, unless
one shifts by a whole number of lattice units.  While the snapshot of
the water shown in the figure has no symmetries, water as a phase
has complete rotational and translational symmetry.

One of the standard tricks to see if two materials differ by a symmetry
is to try to change one into the other smoothly.  Oil and water won't mix,
but I think oil and alcohol do, and alcohol and water certainly do.
By slowly adding more alcohol to oil, and then more water to the alcohol,
one can smoothly interpolate between the two phases.  If they had different
symmetries, there must be a first point when mixing them when the
symmetry changes, and  it is usually easy to tell when that phase
transition happens.

\section{Define the Order Parameter}

Particle physics and condensed--matter physics have quite different
philosophies.  Particle physicists are constantly looking for the building
blocks.  Once pions and protons were discovered to be made of quarks,
they became demoted into engineering problems.  Now that
quarks and electrons and photons are made of strings, and strings are
hard to study (at least experimentally), there is great
anguish in the high--energy community.  Condensed--matter physicists,
on the other hand,
try to understand why messy combinations of zillions of electrons and
nuclei do such interesting simple things.  To them, the fundamental
question is not discovering the underlying quantum mechanical laws,
but in understanding and explaining the new laws that emerge
when many particles interact.

As one might guess, we don't keep track of all the electrons and
protons.\footnote{The particle physicists use order parameter
fields too.  Their order parameter fields also hide lots of details
about what their quarks and gluons are composed of.  The main difference
is that they don't know of what their fields are composed.  It ought
to be reassuring to them that we don't always find our greater knowledge
very helpful.}
We're always looking for the important variables, the important
degrees of freedom.  In a crystal, the important variables are the
motions of the atoms away from their lattice positions.  In a magnet,
the important variable is the local direction of the magnetization
(an arrow pointing to the ``north'' end of the local magnet).
The local magnetization comes from complicated interactions between
the electrons, and is partly due to the little magnets attached to
each electron and partly due to the way the electrons dance around in the
material: these details are for many purposes unimportant.

\begin{figure}[thb]
\epsfxsize=3.5truein
\epsffile{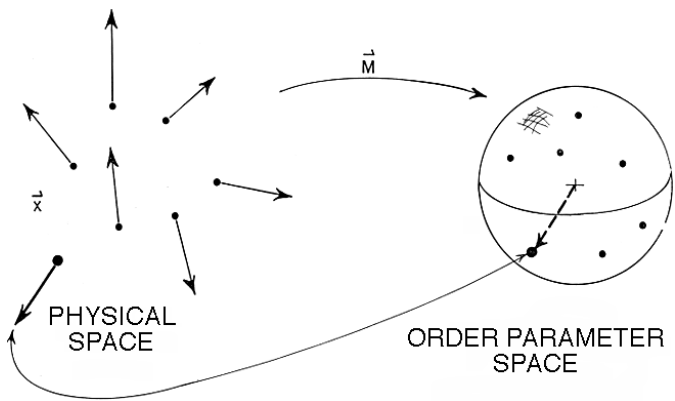}
\caption{{\bf Magnet.}
We take the magnetization $\vec M$ as the order parameter for a magnet.
For a given material at a given temperature, the amount of magnetization
$|\vec M| = M_0$ will be pretty well fixed, but the energy is often pretty
much independent of the direction $\hat M = \vec M / M_0$ of the
magnetization.   (You can think of this as a arrow pointing to the
north end of each atomic magnet.)  Often, the magnetization changes
directions smoothly in different parts of the material.  (That's why
not all pieces of iron are magnetic!)  We describe the current state
of the material by an order parameter field $\vec M({\bf x})$.\hfill\break
The order parameter field is usually thought of as an arrow at each point
in space.  It can also be thought of as a function taking points in
space ${\bf x}$ into points on the sphere $|\vec M| = M_0$.  This sphere
${\cal S}^2$ is the order parameter space for the magnet.
}
\label{fig:4}
\end{figure}

The important variables are combined into an ``order parameter
field''.\footnote{Choosing an order parameter is an art.
Usually it's a new phase which we don't understand yet, and guessing
the order parameter is a piece of figuring out what's going on.  Also,
there is often more than one sensible choice.  In magnets, for example,
one can treat $\vec M$ as a fixed--length vector in ${\cal S}^2$, labelling
the different broken symmetry states.  This is the best choice at low
temperatures, where we study the elementary excitations and topological
defects.  For studying the transition from low to high temperatures,
when the magnetization goes to zero, it is better to consider $\vec M$
as a vector of varying length (a vector in ${\cal R}^3$).  Finding the
simplest description for your needs is often the key to the problem.}
In figure~4, we see the order parameter field for a
magnet.\footnote{Most magnets are crystals, which already
have broken the rotational symmetry.  For some ``Heisenberg'' magnets,
the effects of the crystal on the magnetism is small.  Magnets are really
distinguished by the fact that they break time--reversal symmetry: if
you reverse the arrow of time, the magnetization would change direction!}
At each position ${\bf x}=(x,y,z)$ we have a direction for the local
magnetization ${\vec M}({\bf x})$.  The length of ${\vec M}$ is pretty much
fixed by the material, but the direction of the magnetization is undetermined.
By becoming a magnet, this material has broken the rotational symmetry.
The order parameter ${\vec M}$ labels which of the various broken
symmetry directions  the material has chosen.

The order parameter is a field: at each point in our magnet, ${\vec M}({\bf
x})$
tells the local direction of the field near ${\bf x}$.  Why do we do this?
Why would the magnetization point in different directions in different
parts of the magnet?  Usually, the material has lowest energy when the
order parameter field is uniform, when the symmetry is broken in the
same way throughout space.  In practise, though, the material often
doesn't break symmetry uniformly.  Most pieces of iron don't appear magnetic,
simply because the local magnetization points in different directions
at different places.  The magnetization is already there at the atomic
level: to make a magnet, you pound the different domains until they line up.
We'll see in this lecture that most of the interesting behavior we can study
involves the way the order parameter varies in space.

The order parameter field ${\vec M}({\bf x})$ can be usefully visualized in
two different ways.  On the one hand, one can think of a little vector
attached to each point in space.  On the other hand, we can think of
it as a mapping from real space into order parameter space.  That is,
${\vec M}$ is a function which takes different points in the magnet
onto the surface of a sphere (figure~4).  Mathematicians call the
sphere {\cal S}$^2$, because it locally has two dimensions.  (They don't care
what dimension the sphere is embedded in.)

\begin{figure}[thb]
\epsfxsize=2.5truein
\epsffile{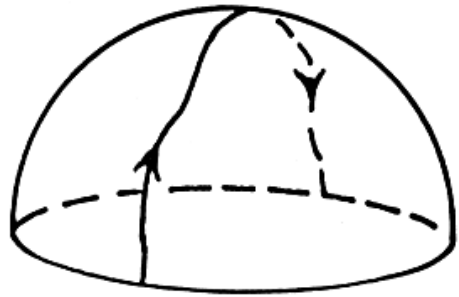}
\caption{{\bf Nematic liquid crystal.}
Nematic liquid crystals are made up of long, thin molecules that prefer
to align with one another.  (Liquid crystal watches are made of nematics.)
Since they don't care much which end is up, their order parameter isn't
precisely the vector $\hat n$ along the axis of the molecules.  Rather,
it is a unit vector up to the equivalence $\hat n \equiv - \hat n$.
The order parameter space is a half-sphere, with antipodal points on the
equator
identified.  Thus, for example, the path shown over the top of the
hemisphere is a closed loop: the two intersections with the equator
correspond to the same orientations of the nematic molecules in space.
}
\label{fig:5}
\end{figure}

Before varying our order parameter in space, let's develop a few more
examples.  The liquid crystal in LCD displays (like those in digital
watches) are nematics.  Nematics are made of long, thin molecules which
tend to line up so that their long axes are parallel.  Nematic liquid
crystals, like magnets, break the rotational symmetry.  Unlike magnets,
though, the main interaction isn't to line up the north poles, but to
line up the axes. (Think of the molecules as American footballs: the same
up and down.)  Thus the order parameter isn't a vector $\vec M$ but a
headless vector $\vec n \equiv -\vec n$.  The order
parameter space is a hemisphere, with opposing points along the equator
identified (figure~5).  This space is called ${\cal RP}^2$ by the
mathematicians (the projective plane), for obscure reasons.

\begin{figure}[thb]
\epsfxsize=2.5truein
\epsffile{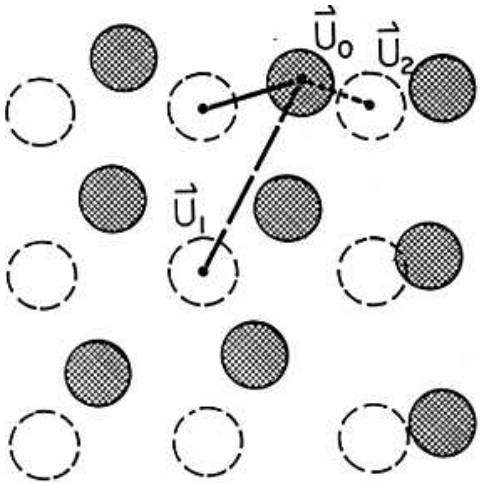}
\caption{ {\bf Two dimensional crystal.}
A crystal consists atoms arranged in regular, repeating rows and columns.
At high temperatures, or when the crystal is deformed or defective, the
atoms will be displaced from their lattice positions.  The displacements
$\vec u$ are shown.  Even better, one can think of $u({\bf x})$ as the
local translation needed to bring the ideal lattice into registry with
atoms in the local neighborhood of $\bf x$.\hfill\break
Also shown is the ambiguity in the definition of $u$.  Which ``ideal'' atom
should we identify with a given ``real'' one?  This ambiguity makes the
order parameter $u$ equivalent to $u+m a \hat x + n a \hat y$.  Instead
of a vector in two dimensional space, the order parameter space is a
square with periodic boundary conditions.
}
\label{fig:6}
\end{figure}

For a crystal, the important degrees of freedom are associated with the
broken translational order.  Consider a two-dimensional crystal which has
lowest energy when in a square lattice, but which is deformed away from
that configuration (figure~6).  This deformation is described by
an arrow connecting the undeformed ideal lattice points with the actual
positions of the atoms.  If we are a bit more careful, we say that
$\vec u({\bf x})$ is that displacement needed to align the ideal lattice in
the local region onto
the real one.  By saying it this way, $\vec u$ is also defined between the
lattice positions: there still is a best displacement which locally lines
up the two lattices.

\begin{figure}[thb]
\epsfxsize=3.5truein
\epsffile{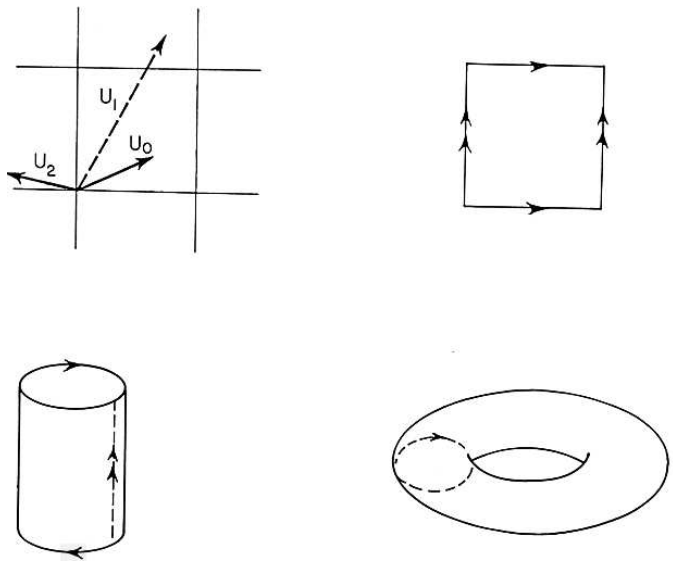}
\caption{{\bf Order parameter space for a two-dimensional crystal.}
Here we see that a square with periodic boundary conditions is a torus.
(A torus is a surface of a doughnut, inner tube, or bagel, depending
on your background.)
}
\label{fig:7}
\end{figure}

The order parameter $\vec u$ isn't really a vector: there is a subtlety.
In general, which ideal atom you associate with a given real one is
ambiguous.  As shown in figure~6, the displacement vector $\vec u$ changes
by a multiple of the lattice constant $a$ when we choose a different
reference atom:
\begin{equation}
\label{eq:equiv}
 \vec u \equiv \vec u + a \hat x = \vec u + m a \hat x + n a \hat y.
\end{equation}
The set of distinct order parameters forms a square with periodic boundary
conditions.  As figure~7 shows, a square with periodic boundary conditions
has the same topology as a torus, {\cal T}$^2$.  (The torus is the surface of a
doughnut, bagel, or inner tube.)

Finally, let's mention that guessing the order parameter (or the broken
symmetry) isn't always so straightforward.  For example, it took many years
before anyone figured out that the order parameter for superconductors
and superfluid Helium 4 is a complex number $\psi$.  The order parameter
field $\psi({\bf x})$ represents the ``condensate wave function'', which
(extremely loosely) is a single quantum state occupied by a large fraction
of the Cooper pairs or helium atoms in the material.  The corresponding
broken symmetry is closely related to the number of particles.  In
``symmetric'', normal liquid helium,
the local number of atoms is conserved: in superfluid helium, the local number
of atoms becomes indeterminate!  (This is because many of the atoms are
condensed into that delocalized wave function.)  Anyhow, the magnitude
of the complex number $\psi$ is a fixed function of temperature, so the
order parameter space is the set of complex numbers of magnitude $|\psi|$.
Thus the order parameter space for superconductors and superfluids is
a circle {\cal S}$^1$.

Now we examine small deformations away from a uniform order parameter field.

\section{Examine the Elementary Excitations}

Its amazing how slow human beings are.  The atoms inside your eyelash
collide with one another a million million times during each time you
blink your eye.  It's not surprising, then, that we spend most of our
time in condensed--matter physics studying those things in materials
that happen slowly.  Typically only vast conspiracies of immense numbers
of atoms can produce the slow behavior that humans can perceive.

\begin{figure}[thb]
\epsfxsize=2.5truein
\epsffile{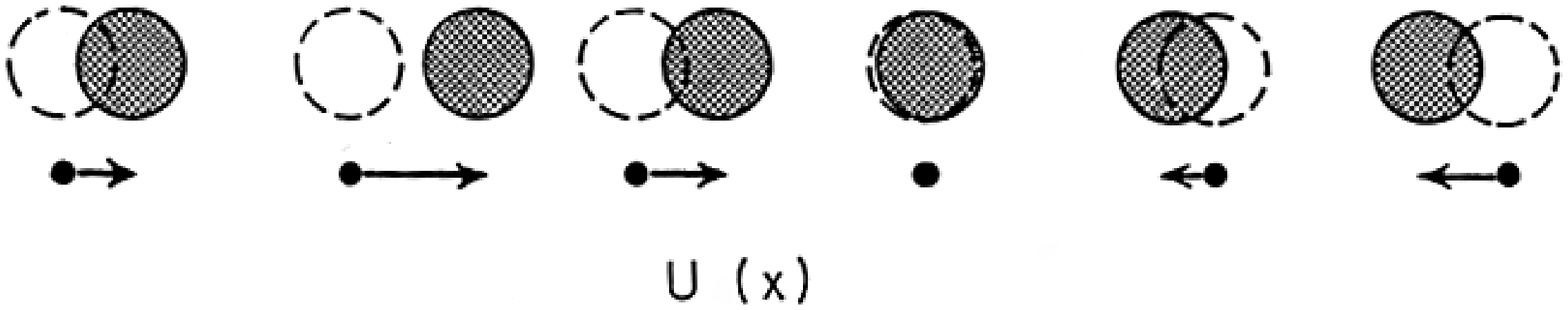}
\caption{{\bf One dimensional crystal: phonons.}
The order parameter field for a one--dimensional crystal is the local
displacement $u(x)$.  Long--wavelength waves in $u(x)$ have low frequencies,
and cause sound.\hfill\break
Crystals are rigid because of the broken translational symmetry.  Because
they are rigid, they fight displacements.  Because there is an underlying
translational symmetry, a uniform displacement costs no energy.  A nearly
uniform displacement, thus, will cost little energy, and thus will have
a low frequency.  These low--frequency elementary excitations are the
sound waves in crystals.
}
\label{fig:8}
\end{figure}

A good example is given by sound waves.  We won't talk about sound waves
in air: air doesn't have any broken symmetries, so it doesn't belong in this
lecture.\footnote{We argue here that low frequency excitations
come from spontaneously broken symmetries.  They can also come from
conserved quantities: since air cannot be created or destroyed, a
long--wavelength density wave cannot relax quickly.}
Consider instead sound in the one-dimensional crystal shown in figure~8.
We describe the material with an order parameter field $u(x)$, where
here $x$ is the position within the material and $x - u(x)$ is the
position of the reference atom within the ideal crystal.

Now, there must be an energy cost for deforming the ideal crystal.
There won't be any cost, though, for a uniform translation: $u(x)\equiv u_0$
has the same energy as the ideal crystal.  (Shoving all the atoms to the
right doesn't cost any energy.)  So, the energy will depend only on
derivatives of the function $u(x)$.  The simplest energy that one
can write looks like
\begin{equation}
\label{eq:Energy}
{\cal E} = \int dx\,(\kappa/2) (du/dx)^2.
\end{equation}
(Higher derivatives won't be important for the low frequencies that
humans can hear.)  Now, you may remember Newton's law $F=m a$.  The
force here is given by the derivative of the energy $F=-(d{\cal E}/du)$.
The mass is represented by the density of the material $\rho$.
Working out the math (a variational derivative and an integration by
parts, for those who are interested) gives us the equation
\begin{equation}
\label{eq:motion}
\rho \ddot u = \kappa (d^2u/dx^2).
\end{equation}
The solutions to this equation
\begin{equation}
\label{eq:phonon}
u(x,t) = u_0 \cos(2 \pi ( x/\lambda -  \nu_\lambda t ))
\end{equation}
represent phonons or sound waves.  The wavelength of the sound
waves is $\lambda$, and the frequency is $\nu_\lambda$.  Plugging
\ref{eq:phonon} into \ref{eq:motion} gives us the relation
\begin{equation}
\label{eq:dispersion}
\nu_\lambda = \sqrt{\kappa/\rho} / \lambda.
\end{equation}

The frequency gets small only when the wavelength gets large.  This
is the vast conspiracy: only huge sloshings of many atoms can happen
slowly.  {\sl Why does the frequency get small?}  Well, there is no
cost to a uniform translation, which is what \ref{eq:phonon} looks like
for infinite wavelength. {\sl Why is there no energy cost for
a uniform displacement?}  Well, there is a translational symmetry:
moving all the atoms the same amount doesn't change their interactions.
{\sl But haven't we broken that symmetry?}  That is precisely the point.

\begin{figure}[thb]
\epsfxsize=2.5truein
\epsffile{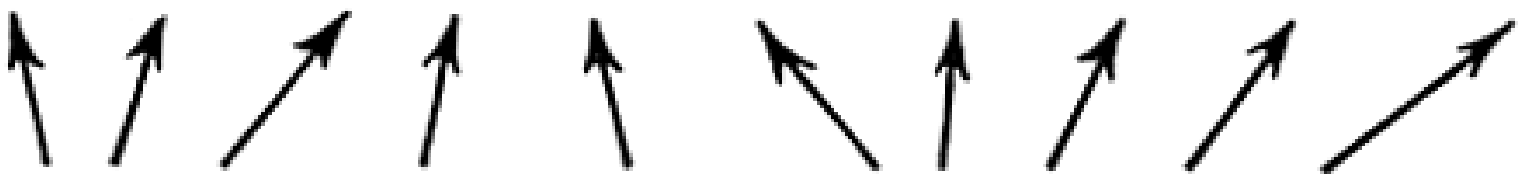}
\epsfxsize=2.5truein
\epsffile{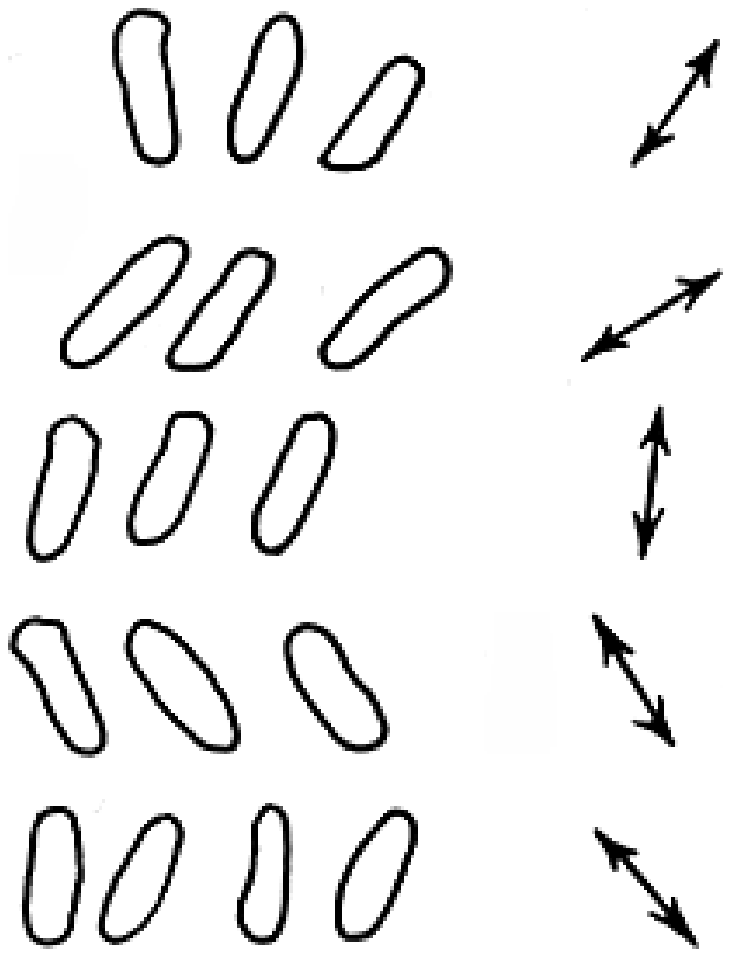}
\caption{{\bf (a)~Magnets: spin waves.}
Magnets break the rotational invariance of space.  Because they resist
twisting the magnetization locally, but don't resist a uniform
twist, they have low energy spin wave excitations.\hfill\break
{\bf (b)~Nematic liquid crystals: rotational waves.}
Nematic liquid crystals also have low--frequency rotational waves.
}
\label{fig:9}
\end{figure}

Long after phonons were understood, Jeffrey Goldstone started to think about
broken symmetries and order parameters in the abstract.  He found a rather
general argument that, whenever a continuous symmetry (rotations, translations,
$SU(3)$, ...) is broken, long--wavelength modulations in the symmetry
direction should have low frequencies.  The fact that the lowest
energy state has a broken symmetry means that the system is stiff:
modulating the order parameter will cost an energy rather like that
in equation \ref{eq:Energy}.  In crystals, the broken translational
order introduces a rigidity to shear deformations, and low frequency
phonons (figure~8).  In magnets, the broken rotational symmetry
leads to a magnetic stiffness and spin waves (figure~9a).
In nematic liquid crystals, the broken rotational symmetry introduces
an orientational elastic stiffness (it pours, but resists bending!)
and rotational waves (figure~9b).

In superfluids, the broken gauge symmetry leads to a stiffness which
results in the superfluidity.  Superfluidity and superconductivity
really aren't any more amazing than the rigidity of solids.  Isn't it
amazing that chairs are rigid?  Push on a few atoms on one side, and
$10^9$ atoms away atoms will move in lock--step.  In the same way,
decreasing the flow in a superfluid must involve a cooperative change
in a macroscopic number of atoms, and thus never happens spontaneously
any more than two parts of the chair ever drift apart.

The low--frequency
Goldstone modes in superfluids are heat waves!  (Don't be jealous:
liquid helium has rather cold heat waves.)  This is often called second
sound, but is really a periodic modulation of the temperature which
passes through the material like sound does through a metal.

O.K., now we're getting the idea.  Just to round things out, what about
superconductors?  They've got a broken gauge symmetry, and
have a stiffness to decays in the superconducting current.  What is the
low energy excitation?  It doesn't have one.  But what about Goldstone's
theorem?  Well, you know about physicists and theorems $\dots$

That's actually quite unfair: Goldstone surely had conditions on his
theorem which excluded superconductors.  Actually, I believe Goldstone
was studying superconductors when he came up with his theorem.  It's just
that everybody forgot the extra conditions, and just remembered that
you always got a low frequency mode when you broke a continuous symmetry.
We of course understood all along why there isn't a Goldstone mode for
superconductors: it's related to the Meissner effect.  The high energy
physicists forgot, though, and had
to rediscover it for themselves.  Now we all call the loophole in
Goldstone's theorem the Higgs mechanism, because (to be truthful)
Higgs and his high--energy friends found a much simpler and more
elegant explanation than we had.  We'll discuss Meissner effects and
the Higgs mechanism in the next lecture.

I'd like to end this section, though, by bringing up another exception
to Goldstone's theorem: one we've known about even longer, but which we
don't have a nice explanation for.  What about the orientational order
in crystals?  Crystals break both the continuous translational order
and the continuous orientational order.  The phonons are the Goldstone
modes for the translations, but {\it there are no orientational Goldstone
modes.}\footnote{In two dimensions, crystals provide another loophole in a 
well-known result, known as the Mermin-Wagner theorem. Hohenberg, Mermin, 
and Wagner, in a series of papers, proved in the 1960's that 
two-dimensional systems with a continuous symmetry cannot have a 
broken symmetry at finite temperature. At least, that's the English 
phrase everyone quotes when they discuss the theorem: they actually 
prove it for several particular systems, including superfluids, 
superconductors, magnets, and translational order in crystals. 
Indeed, crystals in two dimensions do not break the translational 
symmetry: at finite temperatures, the atoms wiggle enough so that 
the atoms don't sit in lock-step over infinite distances (their 
translational correlations decay slowly with distance). But the 
crystals do have a broken orientational symmetry: the crystal 
axes point in the same directions throughout space. (Mermin 
discusses this point in his paper on crystals.) The residual 
translational correlations (the local alignment into rows and 
columns of atoms) introduce long-range forces which force the 
crystalline axes to align, breaking the continuous rotational 
symmetry. Mermin, Wagner, and Hohenberg's methods apply very 
generally, but are not general enough to apply to this case 
(for good reason!)} We'll discuss this further
in the next lecture, but I think this is one of the most interesting
unsolved basic questions in the subject.

\section{Classify the Topological Defects}

\begin{figure}[thb]
\epsfxsize=2.5truein
\epsffile{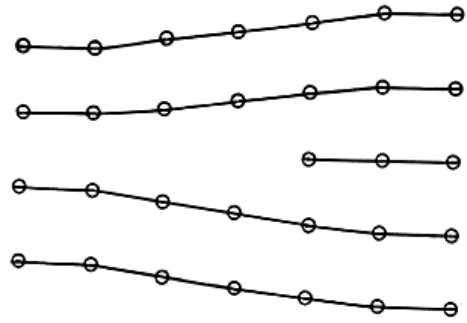}
\caption{{\bf Dislocation in a crystal.}
Here is a topological defect in a crystal.  We can see that one of
the rows of atoms on the right disappears halfway through our sample.
The place where it disappears is a defect, because it doesn't locally
look like a piece of the perfect crystal.  It is a topological
defect because it can't be fixed by any local rearrangement.  No
reshuffling of atoms in the middle of the sample can change the fact
that five rows enter from the right, and only four leave from the left!
\hfill\break
The Burger's vector of a dislocation is the net number of extra rows
and columns, combined into a vector (columns, rows).
}
\label{fig:10}
\end{figure}

When I was in graduate school, the big fashion was topological defects.
Everybody was studying homotopy groups, and finding exotic systems to
write papers about.  It was, in the end, a reasonable thing to
do.\footnote{The next fashion, catastrophe theory, never became
important for anything.}  It is true that
in a typical application you'll be able to figure out what the defects
are without homotopy theory.  You'll spend forever drawing pictures to
convince anyone else, though.  Most important, homotopy theory
helps you to think about defects.

A defect is a tear in the order parameter field.  A topological defect
is a tear that can't be patched.  Consider the piece of 2-D crystal
shown in figure~10.  Starting in the middle of the region shown, there
is an extra row of atoms.  (This is called a dislocation.)
Away from the middle, the crystal locally
looks fine: it's a little distorted, but there is no problem seeing the
square grid and defining an order parameter.  Can we rearrange the atoms
in a small region around the start of the extra row, and patch the defect?

No.  The problem is that we can tell there is an extra row without
ever coming near to the center.  The traditional way of doing this is
to traverse a large loop surrounding the defect, and count the net number of
rows crossed on the path.  In the path shown, there are two rows going
up and three going down: no matter how far we stay from the center, there
will naturally always be an extra row on the right.

\begin{figure}[thb]
\epsfxsize=3.5truein
\epsffile{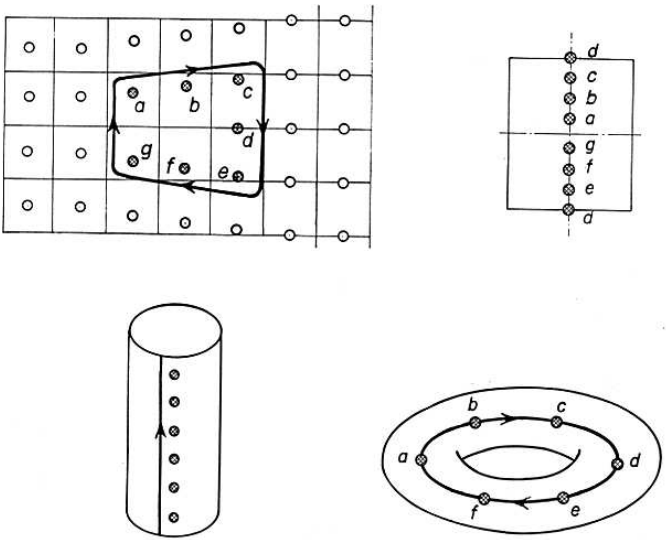}
\caption{{\bf Loop around the dislocation mapped onto order
parameter space.}
How do we think about our defect in terms of order parameters and order
parameter spaces?  Consider a closed loop around the defect.  The order
parameter field $u$ changes as we move around the loop.  The positions of
the atoms around the loop with respect to their local ``ideal'' lattice
drifts upward continuously as we traverse the loop.  This precisely
corresponds to a loop around the order parameter space: the loop passes
once through the hole in the torus.  A loop {\it around} the hole corresponds
to an extra column of atoms.\hfill\break
Moving the atoms slightly will deform the loop, but won't change the number
of times the loop winds through or around the hole.  Two loops which
traverse the torus the same number of times through and around are
equivalent.  The equivalence classes are labelled precisely by pairs
of integers (just like the Burger's vectors), and the first homotopy group of
the torus is ${\cal Z}\times{\cal Z}$.
}
\label{fig:11}
\end{figure}

How can we generalize this basic idea to a general problem with a broken
symmetry?  Remember that the order parameter space for the 2-D square
crystal is a torus (see figure~7).  Remember that the order parameter at
a point is that translation which aligns a perfect square grid to the
deformed grid at that point.  Now, what is the order parameter far to the
left of the defect (a), compared to the value far to the right (d)?  Clearly,
the lattice to the right is shifted vertically by half a lattice constant:
the order parameter has been shifted halfway around the torus.  As shown in
figure~11, along the top half of a clockwise loop the order parameter (position
of the atom within the unit cell) moves upward, and along the bottom
half, again moves upward.  All in all, the order parameter circles once
around the torus.  The winding number around the torus is the net
number of times the torus is circumnavigated when the defect is orbited
once.

This is why they are called topological defects.  Topology is the study
of curves and surfaces where bending and twisting is ignored.  An order
parameter field, no matter how contorted, which doesn't wind around the
torus can always be smoothly bent and twisted back into a uniform state.
If along any loop, though, the order parameter winds either around the
hole or through it a net number of times, then enclosed in that loop
is a defect which cannot be bent or twisted flat: the winding number
can't change by an integer in a smooth and continuous fashion.

How do we categorize the defects for 2-D square crystals?  Well, there
are two integers: the number of times we go around the central hole,
and the number of times we pass through it.  In the traditional
description, this corresponds precisely to the number of extra rows
and columns of atoms we pass by.  This was called the Burger's
vector in the old days, and nobody needed to learn about tori to understand
it.  We now call it the first Homotopy group of the torus:
\begin{equation}
\label{eq:Homotopy}
\Pi_1({\cal T}^2) = {\cal Z} \times {\cal Z}
\end{equation}
where ${\cal Z}$ represents the integers.  That is, a defect is labeled
by two integers $(m,n)$, where $m$ represents the number of extra rows
of atoms on the right-hand part of the loop, and $n$ represents the number
of extra columns of atoms on the bottom.

\begin{figure}[thb]
\epsfxsize=2.5truein
\epsffile{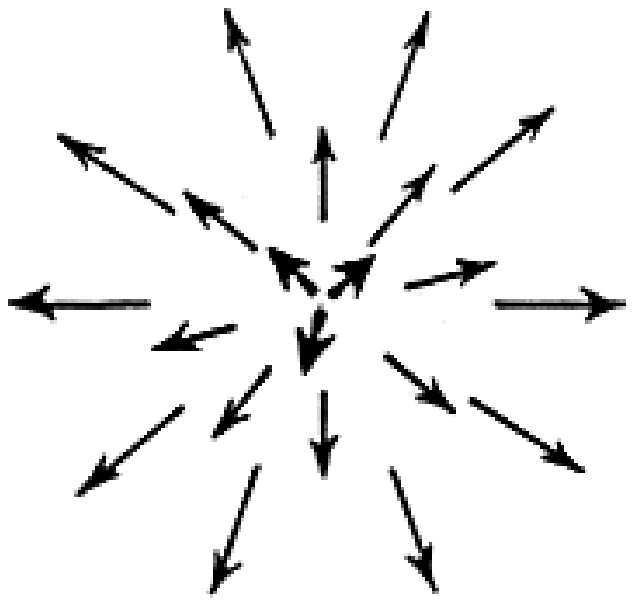}
\epsfxsize=2.5truein
\epsffile{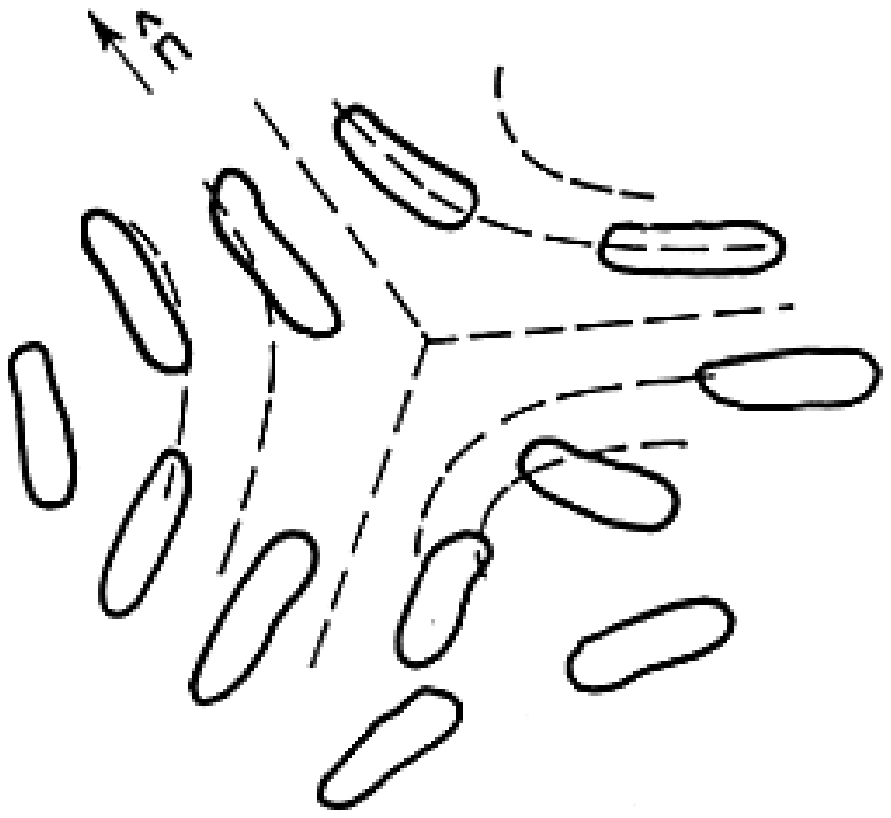}
\caption{{\bf (a)~Hedgehog defect.}
Magnets have no line defects (you can't lasso a basketball), but do have
point defects.  Here is shown the hedgehog defect,
$\vec M({\bf x})= M_0\, \hat x$.  You can't surround a point defect in
three dimensions with a loop, but you can enclose it in a sphere.
The order parameter space, remember, is also a sphere.  The order parameter
field takes the enclosing sphere and maps it onto the order parameter
space, wrapping it exactly once.  The point defects in magnets are
categorized by this {\it wrapping number}: the second Homotopy group
of the sphere is $\cal Z$, the integers.\hfill\break
{\bf (b)~Defect line in a nematic liquid crystal.}
You can't lasso the sphere, but you can lasso a hemisphere!  Here
is the defect corresponding to the path shown in figure~5.  As you
pass clockwise around the defect line, the order parameter rotates
counterclockwise by $180^\circ$.\hfill\break
This path on figure~5 would actually have wrapped around the
right--hand side of the hemisphere.  Wrapping around the left--hand
side would have produced a defect which rotated clockwise by $180^\circ$.
(Imagine that!)  The path in figure~5 is halfway in between, and
illustrates that these two defects are really not different topologically.
}
\label{fig:12}
\end{figure}

Here's where in the lecture I show the practical importance of topological
defects.  Unfortunately for you, I can't enclose a soft copper tube for you
to play with, the way I do in the lecture.  They're a few cents each,
and machinists on two continents have been quite happy to cut them up
for my demonstrations, but they don't pack well into books.  Anyhow,
most metals and copper in particular exhibits what is called work
hardening.  It's easy to bend the tube, but it's amazingly tough to bend it
back.  The soft original copper is relatively defect--free.
To bend, the crystal has to create lots of line
dislocations, which move around to produce the bending.\footnote{This
again is the mysterious lack of rotational Goldstone modes in crystals.}
The line defects get tangled up, and get in the way of any new defects.
So, when you try to bend the tube back, the metal becomes much stiffer.
Work hardening has had a noticable impact on the popular culture.
The magician effortlessly bends the metal bar, and the strongman can't
straighten it $\dots$  Superman bends the rod into a pair of handcuffs
for the criminals $\dots$

Before we explain why these curves form a group, let's give some more
examples of topological defects and how they can be classified.
Figure 12a shows a ``hedgehog'' defect for a magnet.  The magnetization
simply points straight out from the center in all directions.  How
can we tell that there is a defect, always staying far away?  Since this
is a point defect in three dimensions, we have to surround it with a sphere.
As we move around on this sphere in ordinary space, the order parameter
moves around the order parameter space (which also happens to be a sphere,
of radius $|\vec M|$).  In fact, the order parameter space is covered
exactly once as we surround the defect.  This is called the {\it wrapping
number}, and doesn't change as we wiggle the magnetization in smooth ways.
The point defects of magnets are classified by the wrapping number:
\begin{equation}
\label{eq:Homotopy2}
\Pi_2({\cal S}^2) = {\cal Z}.
\end{equation}
Here, the $2$ subscript says that we're studying the second Homotopy group.
It represents the fact that we are surrounding the
defect with a 2-D spherical surface, rather than the 1-D curve we used
in the crystal.\footnote{The zeroth homotopy group classifies domain
walls.  The third homotopy group, applied to defects in three-dimensional
materials, classifies what the condensed matter people call textures
and the particle people sometimes call skyrmions.  The fourth homotopy
group, applied to defects in space--time path integrals, classifies
types of instantons.}

You might get the impression that a strength 7 defect is really just seven
strength 1 defects, stuffed together.  You'd be quite right: occasionally,
they do bunch up, but usually big ones decompose into small ones.
This doesn't mean, though, that adding two defects always gives a bigger one.
In nematic liquid crystals, two line defects are as good as none!
Magnets didn't have any line defects: a loop in real space never surrounds
something it can't smooth out.  Formally, the first homotopy group of the
sphere is zero: you can't loop a basketball.  For a nematic liquid crystal,
though, the order parameter space was a hemisphere (figure~5).  There is
a loop on the hemisphere in figure~5 that you can't get rid of by
twisting and stretching.  It doesn't look like a loop, but you have to
remember that the two opposing points on the equater really represent
the same nematic orientation.  The corresponding defect has a director
field $n$ which rotates $180^\circ$ as the defect is orbited:
figure~12b shows one typical configuration (called an $s=-1/2$ defect).
Now, if you put two of these defects together, they cancel.  (I can't
draw the pictures, but consider it a challenging exercise in geometric
visualization.)  Nematic line defects add modulo~2, like clock arithmetic
in elementary school:
\begin{equation}
\label{eq:Homotopy_2}
\Pi_1({\cal RP}^2) = {\cal Z}_2.
\end{equation}
Two parallel defects can coalesce and heal, even though each one individually
is stable: each goes halfway around the sphere, and the whole loop
can be shrunk to zero.

\begin{figure}[thb]
\epsfxsize=3.5truein
\epsffile{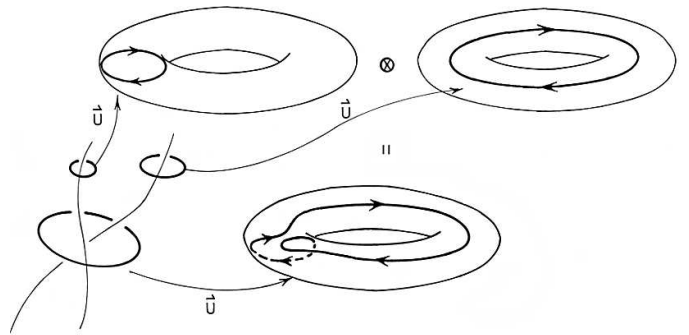}
\caption{{\bf Multiplying two loops.}
The product of two loops is given by starting from their intersection,
traversing the first loop, and then traversing the second.  The inverse
of a loop is clearly the same loop travelled backward: compose the two
and one can shrink them continuously back to nothing.  This definition
makes the homotopy classes into a group.\hfill\break
This multiplication law has a physical interpretation.  If two defect
lines coalesce, their homotopy class must of course be given by the
loop enclosing both.  This large loop can be deformed into two little
loops, so the homotopy class of the coalesced line defect is the product
of the homotopy classes of the individual defects.
}
\label{fig:13}
\end{figure}

Finally, why are these defect categories a group?  A group is a set with
a multiplication law, not necessarily commutative, and an inverse for
each element.  For the first homotopy group, the elements of the group
are equivalence classes of loops: two loops are equivalent if one can
be stretched and twisted onto the other, staying on the manifold at all
times.\footnote{A loop is a continuous mapping from the circle into
the order parameter space: $\theta \rightarrow u(\theta),\>
0\le\theta<2\pi$.  When we encircle the defect with a loop, we get a
loop in order parameter space as shown in figure~4: $\theta \rightarrow
\vec x(\theta)$ is the loop in real space, and $\theta \rightarrow
u(\vec x(\theta))$ is the loop in order parameter space.  Two loops
are equivalent if there is a continuous one-parameter family of loops
connecting one to the other: $u \equiv v$ if there exists
$u_t(\theta)$ continuous both in $\theta$ and in $0\le t\le1$, with
$u_0 \equiv u$ and $u_1 \equiv v$.}
For example, any loop going through the hole from the top (as in the top
right-hand torus in figure~13) is equivalent to any other one.  To multiply
a loop $u$ and a loop $v$, one must first make sure that they meet at some
point (by dragging them together, probably).  Then one defines a new loop
$u \otimes v$ by traversing first the loop $u$ and then
$v$.\footnote{That is,
$u\otimes v(\theta) \equiv u(2\theta)$ for $0\le\theta\le\pi$, and
$\equiv v(2\theta)$ for $\pi \le \theta \le 2\pi$.}

The inverse of a loop $u$ is just the loop which runs along the same
path in the reverse direction.  The identity element consists of the
equivalence class of loops which don't enclose a hole: they can all be
contracted smoothly to a point (and thus to one another).  Finally,
the multiplication law has a direct physical implication: encircling
two defect lines of strength $u$ and $v$ is completely equivalent to
encircling one defect of strength $u\otimes v$.

This all seems pretty trivial: maybe thinking about order parameter spaces
and loops helps one think more clearly, but are there any real uses for
talking about the group structure?  Let me conclude this lecture with an
amazing, physically interesting consequence of the multiplication laws
we described.  There is a fine discussion of this in Mermin's
article\cite{Mermin}, but I learned about it from Dan Stein's
thesis.

\begin{figure}[thb]
\center{\epsfxsize=1truein
\epsffile{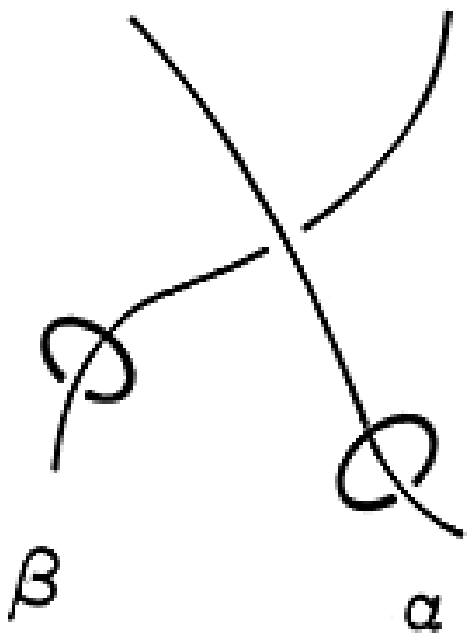}
\hskip 0.2truein
\epsfxsize=1.5truein
\epsffile{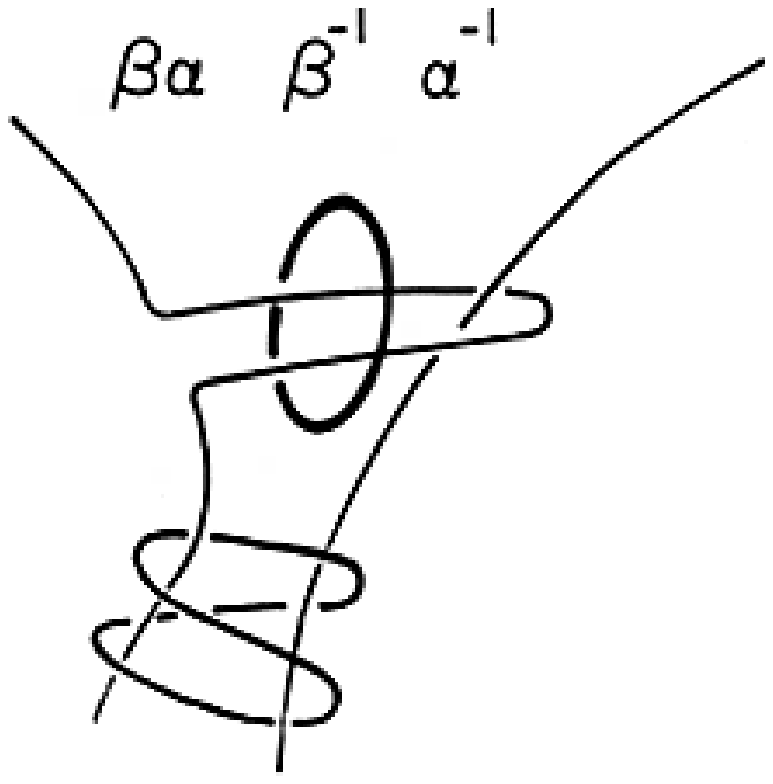}}
\epsfxsize=3.5truein
\epsffile{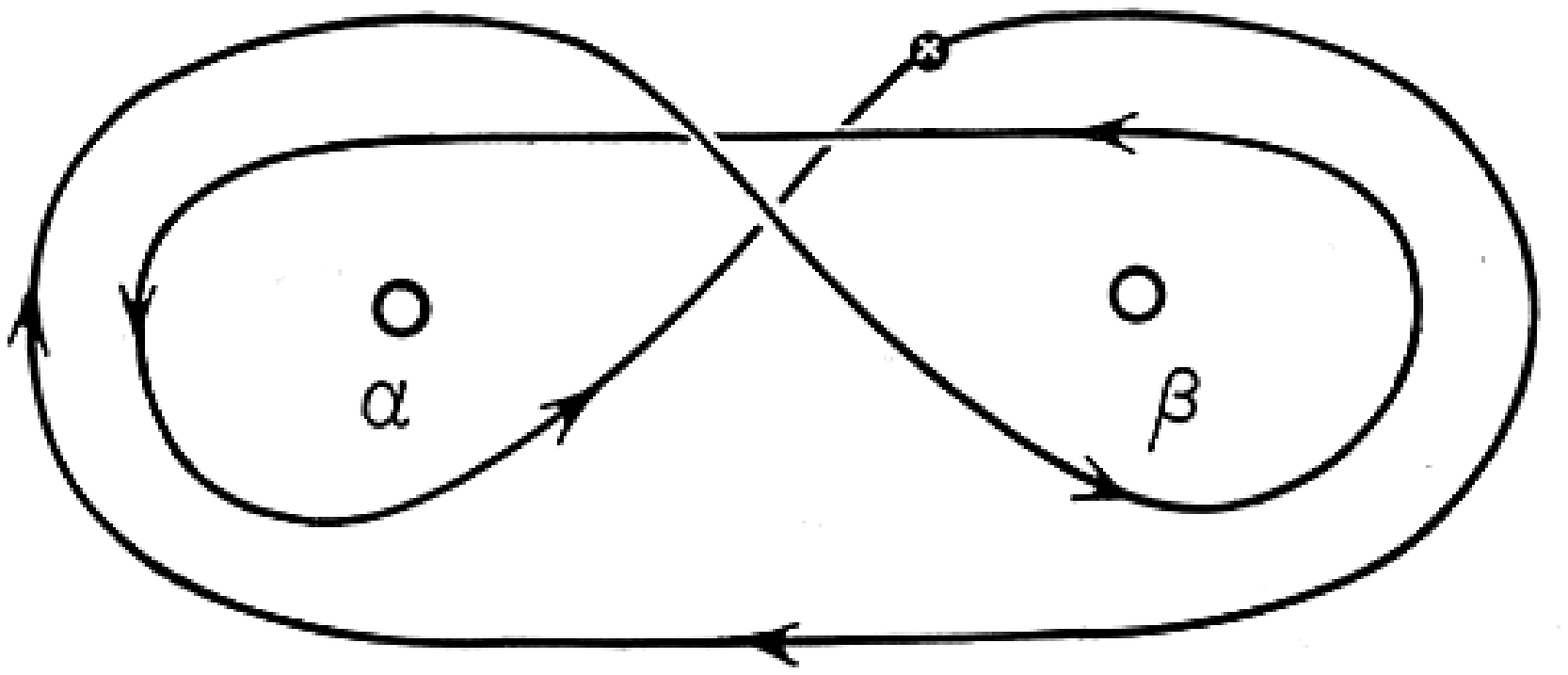}
\caption{{\bf Defect entanglement.}
(a)~Can a defect line of class $\alpha$ pass by a line of class $\beta$,
without
getting topologically entangled?  (b)~We see that we can pass by if we leave
a trail: is the connecting double line topologically trivial?  Encircle the
double line by a loop.  The loop can be wiggled and twisted off the double
line, but it still circles around the two legs of the defects $\alpha$ and
$\beta$.  (c)~The homotopy class of the loop is precisely
$\beta\alpha\beta^{-1}\alpha^{-1}$, which is trivial precisely when
$\beta\alpha = \alpha\beta$.  Thus two defect lines can pass by one another
if their homotopy classes commute!
}
\label{fig:14}
\end{figure}

Can two defect lines cross one another?  Figure~14a shows two defect lines,
of strength (homotopy type) $\alpha$ and $\beta$, which are not parallel.
Suppose there is an external force pulling the $\alpha$ defect past the
$\beta$ one.  Clearly, if we bend and stretch the defect as shown in
figure~14b, it can pass by, but there is a trail left behind, of
two defect lines.  $\alpha$ can really leave $\beta$ behind only if
it is topologically possible to erase the trail.
Can the two lines annihilate one another?  Only if their net strength is
zero, as measured by the loop in 14b.

Now, get two wires and some string.  Bend the wires into the shape found
in figure~14b.  Tie the string into a fairly large loop, surrounding
the doubled portion.  Wiggle the string around, and try to get the
string out from around the doubled section.  You'll find that you can't
completely remove the string, (No fair pulling the string past the cut ends
of the defect lines!) but that you can slide it downward into the
configuration shown in 14c.

Now, in 14c we see that each wire is encircled once clockwise and once
counterclockwise.  Don't they cancel?  Not necessarily!  If you look
carefully, the order of traversal is such that the net homotopy class
is $\beta \alpha \beta^{-1} \alpha^{-1}$, which is only the identity if
$\beta$ and $\alpha$ {\it commute}.  Thus the physical entanglement problem
for defects is directly connected to the group structure of the loops:
commutative defects can pass through one another, noncommutative
defects entangle.

I'd like to be able to tell you that the work hardening in copper is
due to topological entanglements of defects.  It wouldn't be true.
The homotopy group of dislocation lines in fcc copper is commutative.
(It's rather like the 2-D square lattice: if $\alpha = (m,n)$ and
$\beta = (o,p)$ with $m,n,o,p$ the number of extra horizontal and
vertical lines of atoms, then $\alpha \beta = (m+o,n+p) = \beta \alpha$.)
The reason dislocation lines in copper don't pass through one another
is energetic, not topological.  The two dislocation lines interact
strongly with one another, and energetically get stuck when they
try to cross.  Remember at the beginning of the
lecture, I said that there were gaps in the system: the topological
theory can only say when things are impossible to do, not when they are
difficult to do.

I'd like to be able to tell you that this beautiful connection between
the commutativity of the group and the entanglement of defect lines
is nonetheless is important in lots of other contexts.  That too would
not be true.
There are two types of materials I know of which are supposed to suffer from
defect lines which topological entangle.  The first are biaxial nematics,
which were thoroughly analyzed theoretically before anyone found one.
The other are the metallic glasses, where David Nelson has a theory
of defect lines needed to relieve the frustration.  We'll discuss closely
related theories in lecture~3.  Nelson's defects don't commute, and
so can't cross one another.  He originally hoped to explain
the freezing of the metallic glasses into random configurations as an
entanglement of defect lines.  Nobody has ever been able to take this
idea and turn it into a real calculation, though.

Enough, then, of the beautiful and elegant world of homotopy theory:
let's begin to think about what order parameter configurations are
actually formed in practise.

\bigskip\bigskip
\centerline{\bf Acknowledgments}
\bigskip

I'd like to acknowledge NSF grant \# DMR-9118065, and thank NORDITA and the
Technical University of Denmark for their hospitality while these
lectures were written up.

\end{document}

%% file: timeToday.tex
\count60 = \time

\count62 = \count60

\divide\count60 by 60	

\count61 = \count60

\multiply \count61 by -60
\advance \count61 by \count62

\ifnum \count60>12 \global\advance \count60 by -12\fi

\divide\count62 by 60	

\def\timetoday{\ifcase\month\or
  January\or February\or March\or April\or May\or June\or
  July\or August\or September\or October\or November\or December\fi
  \space\number\day, \number\year, 
  \number\count60:\ifnum \count61<10 0\fi\number\count61
  ~\ifnum \count62>12 pm\else am\fi}